\documentclass[aps,pre,reprint,groupedaddress]{revtex4-2}

\usepackage{graphicx}
\usepackage{array,multirow}
\usepackage{amsmath}
\usepackage{color}

\newcommand{\od}[1]{\begingroup\color[rgb]{0,0,0}#1\endgroup}

\begin{document}

\title{Polymer-Chain Configurations in Active and Passive Baths}

\author{Caleb J. Anderson$^{1,2}$, Guillaume Briand$^3$, Olivier Dauchot$^3$, Alberto Fernandez-Nieves$^{1,2,4,5}$}
\affiliation{$^1$Department of Condensed Matter Physics, University of Barcelona, 08028 Barcelona, Spain}
\affiliation{$^2$School of Physics, Georgia Institute of Technology, Atlanta, GA 30332}
\affiliation{$^3$Laboratoire Gulliver, UMR CNRS 7083, ESPCI Paris, PSL University, 10, rue Vauquelin 75231 Paris de cedex 05, France}
\affiliation{$^4$ICREA-Institució Catalana de Recerca i Estudis Avançats, 08010 Barcelona, Spain}
\affiliation{$^5$Institute for Complex Systems (UBICS), University of Barcelona, 08028 Barcelona, Spain}

\date{\today}

\begin{abstract}
The configurations taken by polymers embedded in out-of-equilibrium baths may have broad effects in a variety of biological systems.
As such, they have attracted considerable interest, particularly in simulation studies.
Here we analyze the distribution of configurations taken by a passive flexible chain in a bath of hard, self-propelled, vibrated disks and systematically compare it to that of the same flexible chain in a bath of hard, thermal-like, vibrated disks.
We demonstrate experimentally that the mean length and mean radius of gyration of both chains obey Flory's Law. 
However, the Kuhn length associated with the number of correlated monomers is smaller in the case of the active bath, corresponding to a higher effective temperature.
Importantly, the active bath does not just simply map on a hot equilibrium bath.
Close examination of the chains' configurations indicates a marked bias, with the chain in the active bath more likely assuming configurations with a single prominent bend.
\end{abstract}

\maketitle

\section{Introduction}

A linear polymer in a good solvent at equilibrium has conformations that correspond to a self-avoiding random walk with a step size equal to the Kuhn length, \(b\), defined as the length between monomers for orientational correlations to be lost.
The Kuhn length is directly related to the bending stiffness of the polymer, \(\kappa\), and depends inversely on the thermal energy: \(b = \frac{2 \kappa}{k_B T}.\) 
In terms of \(b\), the average length-scales of the polymer, such as the average end-to-end distance, \(\langle R_{ee}\rangle \), and the average radius gyration, \(\langle R_g\rangle \), scale as  $b \cdot N_b^\nu$, where \(N_b\) is the length of the polymer, \(L\), divided by the Kuhn length, \(N_b=L/b\), and \(\nu\) is a scaling exponent called the Flory exponent.
For a self-avoiding random walk in 2D, \(\nu=3/4\) \cite{Polymer2003,deGennes1979}. 

Far less is known about the behavior of linear polymers in out-of-equilibrium baths, despite its potential significance in biology, where biopolymers, such as proteins and filaments, are forced out of equilibrium by the presence of motor proteins \cite{Kruse2004,JULICHER2007,Prost2015,Weber2015,Oyama2019}.
To date, most of the work on out-of-equilibrium polymers has focused on two model cases.
The first case is active polymers, which consist of monomers that experience active forces that drive the polymer out of equilibrium; these have been studied both in simulations and experiments \cite{Winkler2020,Locatelli2021}.
The second case pertains to passive linear polymers in an out-of-equilibrium bath, comprised, for example, of active particles; these have been solely studied by computer simulations \cite{Kaiser2014, Shin2015, Harder2014,Xia2019,Cao2020}.
For \(\kappa=0\), the polymer was found to slightly swell with increasing bath activity \cite{Kaiser2014}, while for \(\kappa \neq 0\), the polymer was found to shrink \cite{Shin2015}.

In this paper, we directly test some of the simulation expectations using connected stainless steel ball-chains as a model passive polymer and baths of radially symmetric or asymmetric disks on a vibrated plate. 
While the system composed of symmetric particles is out-of-equilibrium, it has been shown previously \cite{Junot2017,Briand2016,Lanoiselee2018} that the assembly of such particles behaves in a manner close to an equilibrium hard-disk liquid. 
Therefore the symmetric particles can be taken as a model passive bath, while the asymmetric particles, which behave as polar self-propelled particles, constitute the active bath.
We find that the growth of the radius of gyration of the polymer in an active bath with the number of monomers is consistent with what is expected for the polymer in a passive bath.
However, activity results in smaller values of \(R_g\), indicating the polymer shrinks in active baths.
Additionally, the actual polymer chain conformations can be significantly different; in active baths, the polymer adopts significantly more ``hairpin''-like structures than in passive baths.

\section{Methods}

We use the experimental system described in detail in Reference \cite{Deseigne2012} and subject a collection of macroscopic circular grains and a ball-chain to a well-controlled vertical vibration while they are confined to a \(2.4\) mm gap between two horizontal glass plates. 
An electromagnetic servo-controlled shaker (V455/6-PA1000L,LDS) coupled to a triaxial accelerometer (356B18, PCB Electronics) allows producing a sinusoidal vibration in the bottom plate.
The resulting contacts between the grains and the glass plates cause the grains to experience horizontal displacement over time. 
We work at a frequency \(f= 120\) Hz and set the acceleration relative to gravity to \(\Gamma = a \left( 2\pi f\right)^2/g = 2.0\), which corresponds to a peak vertical displacement of $a=34~\mu$m.
An accelerometer is used to ensure that no resonances of the experimental set up are present at this working frequency, that the horizontal to vertical ratio is lower than \(10^{-2}\), and that the spatial homogeneity of the vibrations across the bottom plate is within \(1\%\).

We carry out experiments with embedded chains consisting of various numbers, \(N\), of hollow metal beads with a diameter \(a_0 =(2.30 \pm 0.05)\) mm  and an average center-to-center distance \(\sigma = (3.10 \pm 0.05)\) mm in quasi 2D baths of active or passive disks. 
The active bath is composed of polar, self-propelled grains, which are micro-machined monodisperse copper-beryllium disks with a diameter \( d_{0} = 4\) mm, and an off-center tip and a glued rubber skate located in diametrically opposite positions; these raise the total height of the disks to \(h=2\) mm. 
The two ``legs'' have different mechanical response endowing the particles with a polar axis.
At the working frequency, the disks perform a persistent random walk, with a speed \(v_{0} = (4.59 \pm 0.01)~d_{0}/\text{s}\) and a rotational diffusion constant \(D_{\theta} = 0.76~ \text{rad}^{2}/\text{s}\). 
Together, these two measurements can be combined to obtain a persistence length \(\xi = \pi^{2}v_{0}/(2D_{\theta}) = 14.6~d_{0}\) \cite{Deseigne2012}.     

The passive bath is composed of isotropic grains which are disks made of the same metal, same diameter and same height, but that are rotationally invariant. 
The contact with the vibrating plate results in the disks executing a random walk with diffusion constant \(D= (0.78 \pm 0.01)~d_{0}^{2}/\text{s} \). 

In addition to being confined between the two horizontal glass plates, the disks are constrained horizontally by a flower shaped boundary [see Fig.\ref{fig:Rg}(a)], which frustrates the tendency of active particles to accumulate at flat or concave walls \cite{Berke2008,Li2009,Elgeti2013}. 
The area fraction within the cell boundaries is held constant and equal to \(\Phi=0.088\) throughout the experiments. 

For each trial, we take images with a CCD camera at \(30~\text{fps}\) for \(10\) min and tracked the positions of each link of the polymer chain as a function of time so that a typical trial produced \(18000\) images. We then exclude from our data all images in which any bead of the chain is located within approximately \(5d_0\) from the boundary.

\section{Polymer Shrinking}

To check whether our model polymer chains expand or contract in the active bath relative to the passive bath, we measure their radii of gyration for each time step by first computing the gyration tensor: \(S_{\alpha \beta} = \sum_{i=1}^N \Delta \alpha_i \Delta \beta_i \), where \( \alpha, \beta \in {x,y} \) and \(\Delta \alpha_i\) and \(\Delta \beta_i\) are the \(\alpha\) and \(\beta\) positions of a monomer in the chain relative to the center of mass of the polymer chain; the index \(i\) runs over all monomers. 
The gyration tensor is symmetric and has two real eigenvalues,  \(\lambda_2>\lambda_1\), so that \(\lambda_2\) is the axis corresponding to the maximum \(1\)-dimensional radius of gyration.
The squared radius of gyration is then \(R_g^2= Tr(S) = \lambda_1+\lambda_2\), corresponding to \(R_g^2 = \sum_{i=1}^N (\Delta x_i^2 + \Delta y_i^2)\).

\begin{figure}[!h]
	\centering
	\includegraphics[width=3.2in]{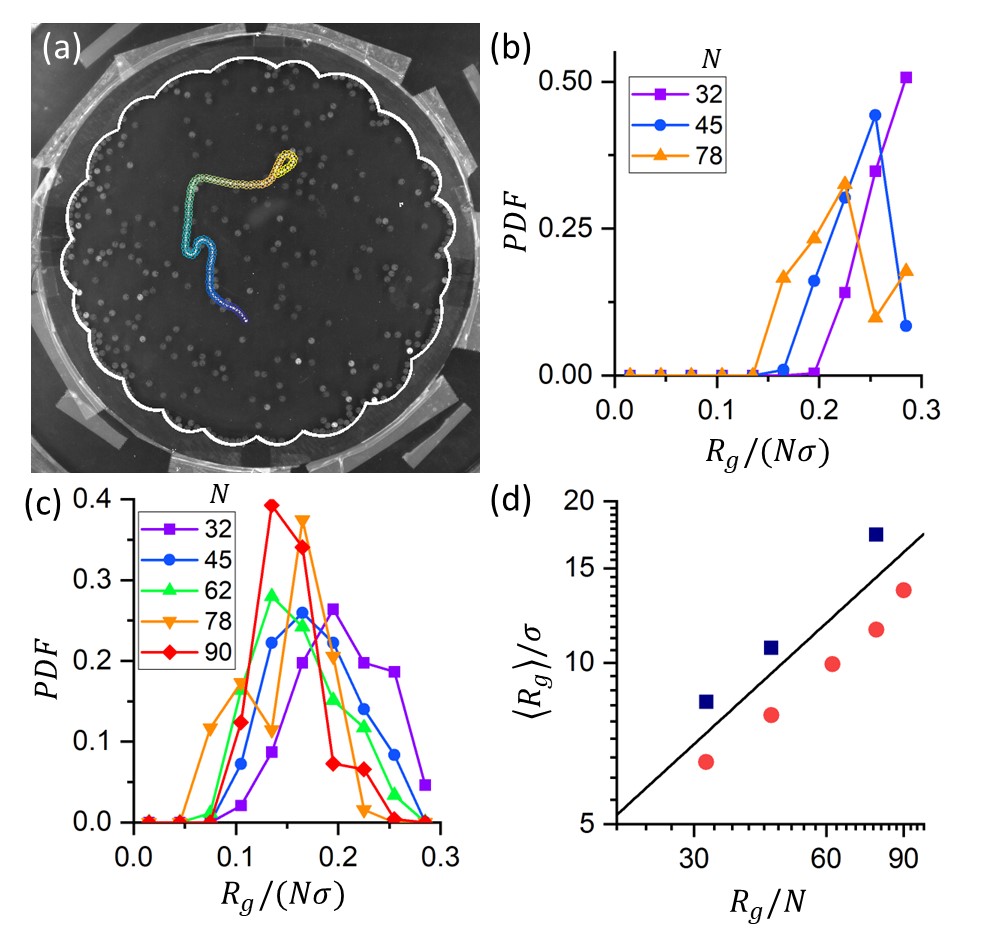}
	\caption{
	(a) A typical image of an experiment with the monomers of the chain tracked in color and the flower shaped cell highlighted in white.  The beads of the chain have been highlighted in colors that range from blue (dark) to yellow (light) from one end of the chain to the other.
	(b) The probability distribution functions for \(R_g /(N\sigma)\) for three different lengths of chains in a passive bath. In these units, higher N would correspond to lower \(\kappa\) in equilibrium systems, which results in lower radii of gyration.
	(c) The probability distribution functions for \(R_g/(N\sigma)\) for  five different lengths of chains immersed in the active bath.
	(d) \(\langle R_g \rangle / \sigma\) for passive (blue squares) and active (red circles) baths. The black line represents \(R_g^{0.75}\), corresponding to a self avoiding random walk.
	}
	\label{fig:Rg}
\end{figure}

Figs. \ref{fig:Rg}(b) and \ref{fig:Rg}(c) show the probability distribution functions of our measured radii of gyration for the polymer chains immersed in both passive and active baths, respectively, with the radius of gyration scaled by \(N \sigma\).
With this scaling, the maximum end-to-end length of the polymer is \(R_{ee}/(N \sigma) = 1\), corresponding to a maximum radius of gyration equivalent to that of an infinitesimally thin rod: \(\frac{R_g}{N \sigma} = \left(\int_{-1/2}^{1/2} x^2 dx \right)^\frac{1}{2}= \sqrt{1/12} \approx 0.29.\)

In a bath of passive particles, the shortest chain, with \(N=32\) monomers, acts like a very stiff polymer, with \od{the most probable value of} \(R_g\) near the maximum limit; see red squares in Fig. \ref{fig:Rg}(b).
In this case, the chain is almost always fully extended.
As \(N\) increases, the relative radius of gyration \(R_g/(N\sigma)\) begins to shrink.
This is because polymer chains with normalized length \(L/(N\sigma)=1\) contains more Kuhn lengths; we are thus increasing the polymer length relative to its persistence length.

When we instead immerse the polymer chains in the active bath, we find that they posses a much lower \(R_g/(N\sigma)\) for a given length [see Fig. \ref{fig:Rg}(c)], indicating that to leading order, the chains effectively have a decreased Kuhn length in the active bath, making it more flexible; this is in agreement with the simulation predictions in Refs. \cite{Shin2015, Kaiser2014}.
 
We summarize these results in Fig. \ref{fig:Rg}(d), where we show that both the chain immersed in the passive bath and the chain immersed in the active bath, both follow the expected Flory Law \(R_g\propto N^{3/4}\), at least within the  contour lengths we are able to study.
This is consistent with computer simulations \cite{Kaiser2014}.
Fig. \ref{fig:Rg}(d) also shows that the average \(R_g\) for the chain immersed in the active bath is always smaller (red circles) compared to the passive bath (navy squares), consistent with the chain in the active bath having a shorter Kuhn length.
Because the physical stiffness of our chain has not decreased in reality, the increased flexibility of the chain in this case, can be taken as a sign of an increased effective thermal energy of the bath, \(k_B T_{\text{eff}}\), which is often considered as one of the main effects of activity in low density systems \cite{Palacci2010,Loi2011,Ginot2015,Flenner2016,Caprini2019}.

Overall, our results might seem to suggest that the primary effect of the active bath is to increase the effective temperature.
However, we shall see now that the active bath cannot simply be mapped onto a passive one, as we also find remarkable differences in the typical shapes adopted by the chains in active and passive baths.

\section{Hairpins and Coils}

Simulation work has predicted that, in addition to shrinking, polymers in active baths are much more likely to adopt hairpin configurations \cite{Harder2014}; these contain a single prominent bend and are otherwise extended.
One way to measure the prevalence of this type of configuration is to measure the so called acylindricity of the polymer chain and compare it to its radius of gyration.
The acylindricity is defined as \(A^2 = \frac{\lambda_2 -\lambda_1}{R_g^2}\), and it measures the relative difference between the 1-dimensional radii of gyration.
If these radii are equal, as in the case of a uniform circle or a square, then \(A=0\). Conversely, \(A\) is maximum for a line, which has \(\lambda_1=0\) and \(A=1\).

\begin{figure}[!h]
	\centering
	\includegraphics[width=3.2in]{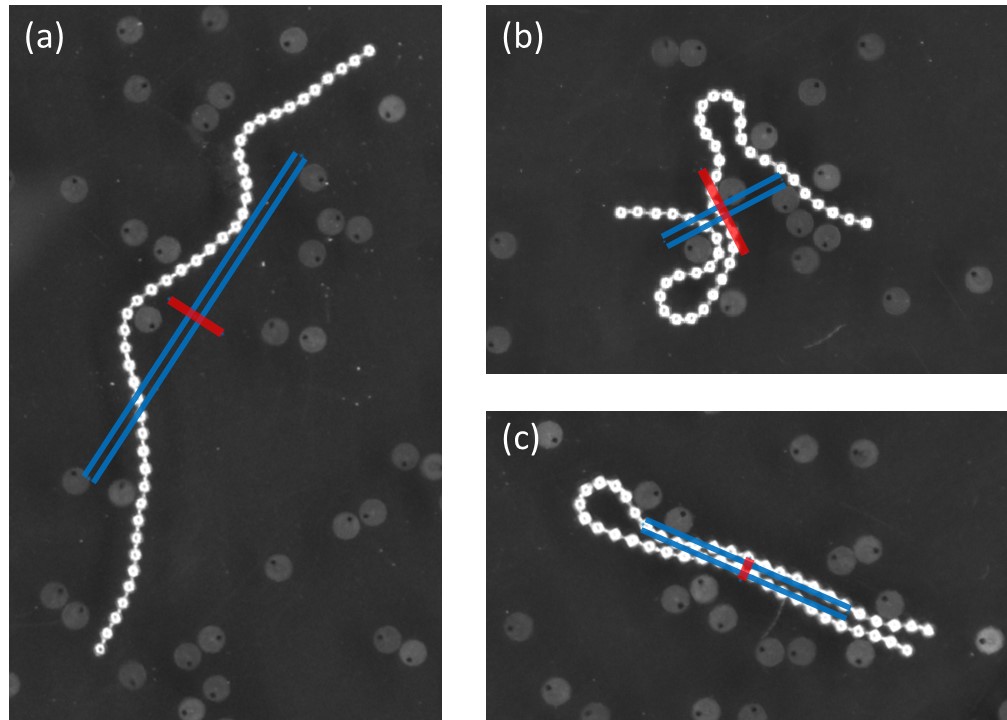}
	\caption{
	(a) A chain with \(N=45\) in a nearly fully extended configuration..
	(b) The chain in a more tightly confined configuration.
	(c) The chain in a hairpin configuration.  The lines represent the principal axes of the configuration and their length is \(2\sqrt{\lambda}\).
	}
	\label{fig:chp6:Hairpins}
\end{figure}

\begin{figure*}[!hbt]
	\centering
	\includegraphics[width=5in]{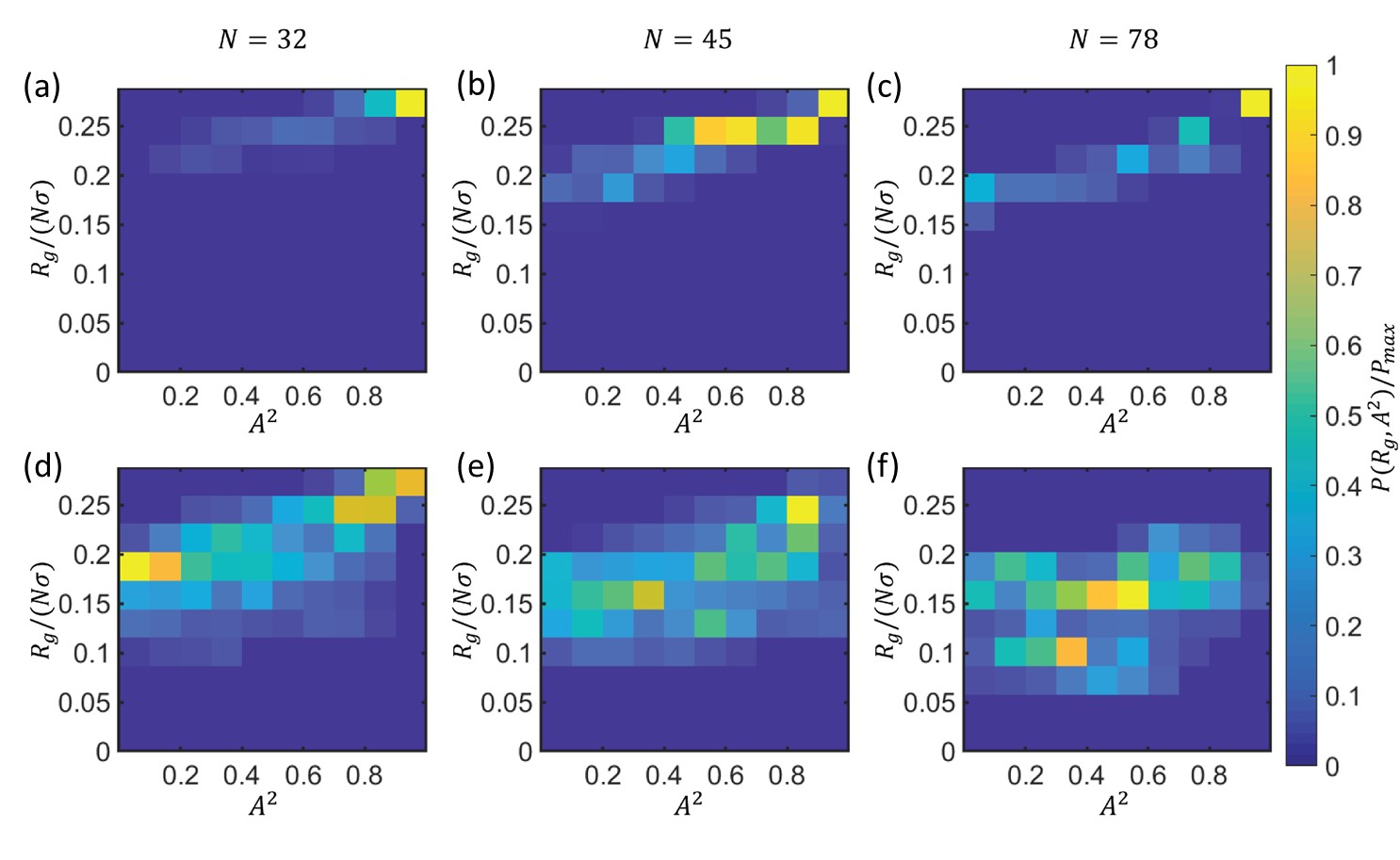}
	\caption{
	(a-c) The measured prevalence of conformations with various \(R_g\) and \(A^2\) for polymers with \(N=32, 45, \text{and } 78\) monomers, respectively, immersed in a passive bath.
	(d-f) The prevalence of conformations for polymers with \(N=32, 45, \text{and } 78\) monomers, respectively, immersed in an active bath.
	}
	\label{fig:chp6:R_A}
\end{figure*}

Simultaneous measurements of \(A^2\) and \(R_g\) thus allow detecting whether our polymer chains adopt hairpin-like configurations.
As an example, consider the three configurations shown in Fig. \ref{fig:chp6:Hairpins}(a-c).
We have added lines that represent the lengths and orientations of their principal radii of gyration to more easily compare them with each other; each double blue line has a length of \(2\sqrt{\lambda_2}\), and each red line has length \(2\sqrt{\lambda_1}\).
The configuration in Fig. \ref{fig:chp6:Hairpins}(a) has \(A^2=0.95\) and \(R_g/(N\sigma)=0.26\), which are near their possible maximum values.
The acylindricity is large because \(\lambda_2 \gg \lambda_1\), and \(R_g\) is also large because the polymer is fully extended.
In comparison, the configuration in Fig. \ref{fig:chp6:Hairpins}(b) is spread out more isotropically, corresponding to \(\lambda_1 \approx \lambda_2\) and a smaller acylindricity \(A^2 = 0.31\). 
At the same time, the chain is more compact, which reduces the radius of gyration to \(R_g/(N\sigma) =0.11\).
Figure \ref{fig:chp6:Hairpins}(c) shows an example of a hairpin configuration with \(A^2 = 0.98\) and \(R_g = 0.14\).
The acylindricities of these configurations are very high, because the hairpins are highly anisotropic, but they all have much lower values of \(R_g\) relative to the values expected for a fully extended chain.

We find that our polymer chains adopt many more hairpin conformations when immersed in active, as compared to passive, baths.
Figs. \ref{fig:chp6:R_A}(a-c) show the probability of conformations in a passive bath in terms of \(R_g\) and \(A^2\), for polymer lengths corresponding to \(N = 32, 45, 78\), respectively.
The gray scale represents the probability of the polymer having the given \(R_g\) and \(A^2\); note we scale the probabilities for each trial by the highest probability in that trial, so that all plots can be shown with the same gray scale.
In passive baths, the radius of gyration and acylindricity of the polymer chain are closely related; there is little spread in the data corresponding to the largest probabilities.
Any reduction in \(R_g\) is thus accompanied by a reduction in \(A^2\).
In contrast, for active baths, there are many more conformations with low \(R_g\) for a given \(A^2\) [Figs. \ref{fig:chp6:R_A}(d-f)], indicating the presence of an appreciable number of hairpin configurations; this agrees with expectations from computer simulations \cite{Harder2014}.

\begin{figure}
	\centering
	\includegraphics[width=3.375 in]{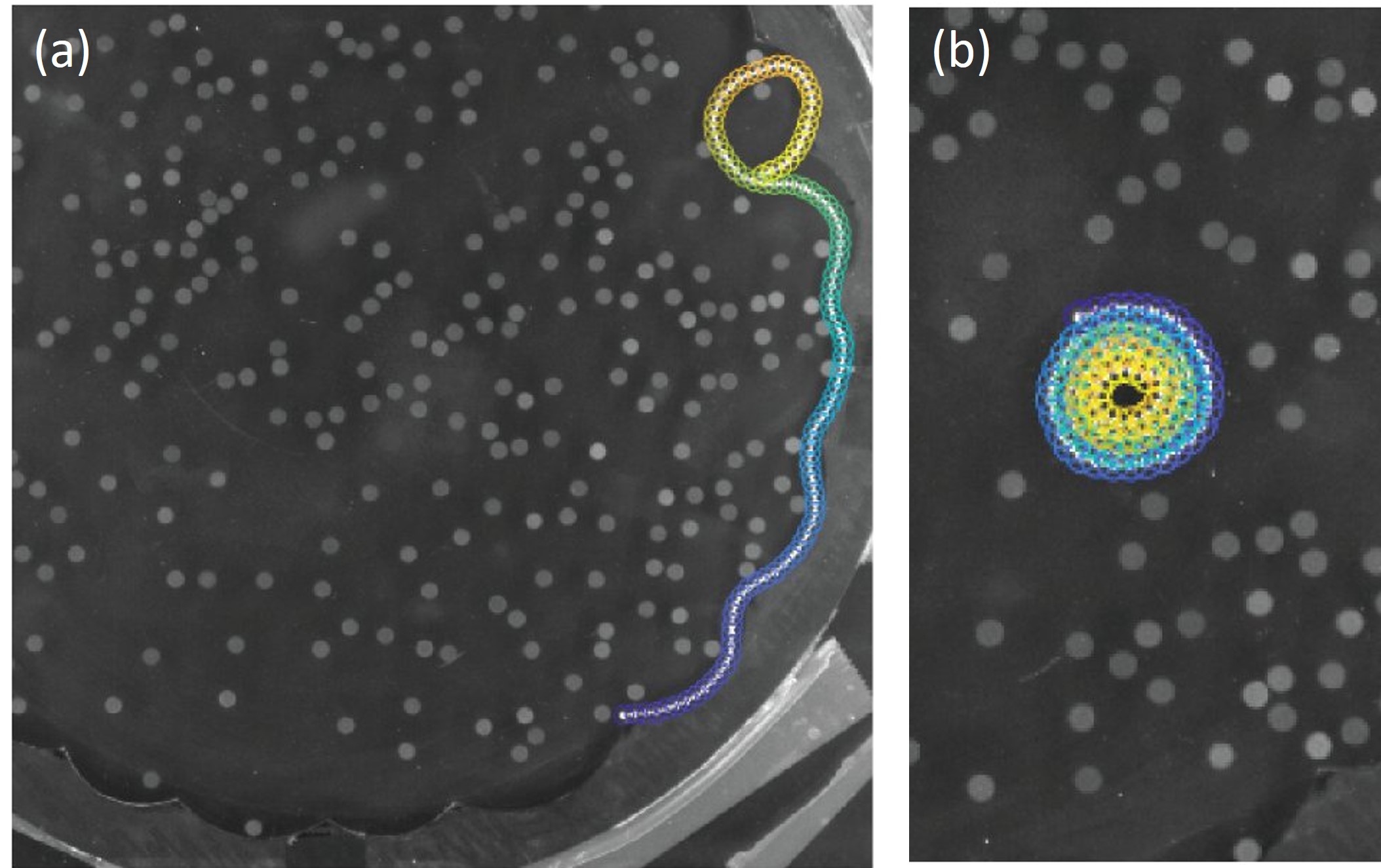}
	\caption{
	(a) An example of a chain in a passive bath depleting to the boundaries. Once this happen, the chain never returns to the bulk.
	(b) An example of a chain coiling in a passive bath. This is also a steady state as the grains cannot exert any force that would uncoil it.
	}
	\label{fig:chp6:Snail}
\end{figure}

We also note that in the presence of passive baths, the polymer chain occasionally reaches a steady state configuration where it is completely depleted to the wall, such as that illustrated for a chain with \(N=100\) monomers in Fig. \ref{fig:chp6:Snail}(a).
This is never observed in the presence of active baths, as we always find active particles near the wall that are able to eventually push the chain back into the bulk of the experimental cell.

In other instances also in passive baths, we find that the polymer collapses into a coil, as illustrated in Fig. \ref{fig:chp6:Snail}(b). 
This coiled configuration is the same as the one found in the simulations in Ref. \cite{Liu2019}.
However, in this case, the polymer is inside a chiral active bath composed of self-propelled particles with a non-zero angular velocity in addition to their average directed motion velocity.
Additional experiments would be needed to asses whether there is some hidden chirality in our set up with passive particles or if the spiral configuration is just stable in two-dimensions when the particle density in the bath is low enough so that no particles from the bath are trapped within a loop configuration of the polymer.

\section{Conclusion}
In this paper, we have used a well-studied two-dimensional system of self propelled particles to explore the configurations of a passive polymer embedded in an active bath.
Our results show that, in both passive and active baths, the average radius of gyration increases with the number of monomers in a manner consistent with the Flory scaling law of equilibrium polymers.
This makes it tempting to compare the effect of the active bath to an escalated effective temperature.
However, by comparing simultaneous measurements of the radii of gyration and acylindricity of the polymers, we verify that the activity of the bath changes the configurations of the polymer, skewing the likely configurations towards those that are more ``hairpin-like'', i.e. configurations with a single prominent bend caused by one or more active particles briefly penetrating and dragging the polymer.
Importantly, in our passive particle bath, the polymer can adopt steady-state configurations not seen in our active particle bath; these configurations correspond to polymer chains that are either depleted to the boundary of the cell, when by chance, there are no particles between the chain and the wall, or to spiral states, when the polymer begins to close on itself with no particles inside the loop.

Our results verify various predictions from simulations of passive polymers in an active bath and may be a step towards further understanding of polymer collapse in a variety of biological situations.

\begin{acknowledgments}
We thank MCIN/AEI/10.13039/501100011033/FEDER,289 (grant No. PID2021-122369NB-100), as well as the FLAMEL (NSF DGE-1258425) and REU (NSF Grant GR10002751) programs, for financial support.
\end{acknowledgments}

\bibliography{references}

\end{document}